\documentclass[fleqn,aps,pra,amsmath,amssymb,preprint,showpacs]{revtex4-1}
\usepackage{amsmath}
\usepackage{graphicx}
\usepackage{dcolumn}
\usepackage{bm}
\usepackage{textcomp}
\usepackage{natbib}

\def\ket#1{\mathinner{|{#1}\rangle}}

\renewcommand\Re{\operatorname{Re}}
\renewcommand\Im{\operatorname{Im}}
\begin{document}
\title{Sub- and super-luminal light propagation using a Rydberg state}
\author{Vineet Bharti}
\author{Vasant Natarajan}
\affiliation{Department of Physics, Indian Institute of Science, Bangalore-560\,012, India}

\begin{abstract}
We present a theoretical study to investigate sub- and super-luminal light propagation in a rubidium atomic system consisting of a Rydberg state by using density matrix formalism. The analysis is performed in a 4-level vee+ladder system interacting with a weak probe, and strong control and switching fields. The dispersion and absorption profiles are shown for stationary atoms as well as for moving atoms by carrying out Doppler averaging at room temperature. We also present the group index variation with control Rabi frequency and observe that a transparent medium can be switched from sub- to super-luminal propagation in the presence of switching field. Finally, the transient response of the medium is discussed, which shows that the considered 4-level scheme has potential applications in absorptive optical switching. 


\end{abstract}

\maketitle

\section{Introduction}
Over the last two decades, the phenomenon of electromagnetically induced transparency (EIT), in which an initially absorbing medium for a weak probe
field becomes nearly transparent in the presence of a strong control field, has been extensively studied in 3-level configurations---lambda ($\Lambda$), vee (V), and ladder ($\Xi$) \cite{BIH91,HAR97,FIM05}. There are parallel developments to demonstrate Rydberg EIT by using the excited level as Rydberg state in ladder scheme. These studies have been utilized for the measurement of fine and hyperfine splitting of Rydberg states in rubidium \cite{MJA07,TNL13}, electro-optic control of atom-light interactions \cite{BMW08}, determination of atom-wall-induced light shifts \cite{KSB09}, and microwave dressing of Rydberg states \cite{TPM11}.

The presence of additional energy levels and laser fields from the usual 3-level EIT systems leads to the modification of transparency window and allows the possibility of electromagnetically induced absorption (EIA)---a phenomenon in which a transparent medium shows enhanced absorption at line center. EIA has been studied---both theoretically as well as experimentally---mainly in 4-level N-type ($\Lambda$+V) systems \cite{GWR04,BMW09,CPN12,BHW13}. Recently, our group has theoretically shown that EIA resonances are also possible in a new kind of 4-level system in vee+ladder configuration (V+$\Xi$) \cite{BHN15}. We extended this study to investigate the wavelength mismatch effect in EIA and reported that EIA resonances can be studied using a Rydberg state excited with diode lasers \cite{BWN16}. 

Just like the imaginary part of the induced coherence on the probe transition can lead to anomalous absorption, the real part can lead to anomalous dispersion. This can be used for applications such as sub- and super-luminal light propagation \cite{HHD99,KSZ99,KRS01,WKD00,DKW01}. Enhanced transmission leads to sub-luminal propagation; therefore an EIT medium can only be used for this. On the other hand, enhanced absorption results in super-luminal propagation, and the presence of four levels in an EIA configuration allows for switching between enhanced transmission and enhanced absorption by varying the strengths of the two control fields. Therefore, an EIA medium can be switched between sub- and super-luminal propagation. This kind of switching between the two has also been studied (theoretically) in N-type systems \cite{AGD04,HGB05,KWK09,KBW14}.

In this work, we present for the first time details of sub- and super-luminal light propagation in a 4-level system \textit{comprising a Rydberg state for the uppermost level}. Rydberg states are important for applications in quantum-information processing, but a scheme involving Rydberg states is not possible in the widely studied N-type systems. Furthermore, in contrast to our work in Ref.~\cite{BWN16} where we study only Rydberg EIA, we concentrate here on the dispersion curve (its slope) and refractive index of the medium. We also present the transient response of our system, which shows that the medium can be used for absorptive optical switching.

\section{Theoretical considerations}
\subsection{Density matrix analysis}
\begin{figure}
	\centering
	\includegraphics[width=.4\textwidth]{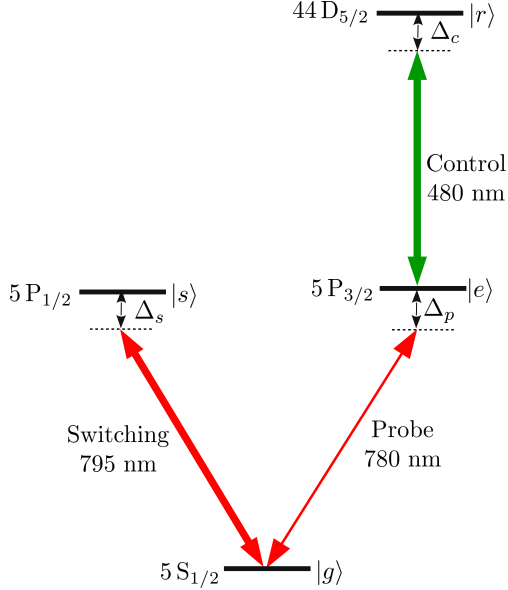}
	\caption{(Color online) 4-level vee + ladder system under consideration.}
	\label{model}
\end{figure}

The 4-level vee+ladder system is shown in Fig.\ \ref{model}. The ground state $\ket{g}$  is coupled with state $\ket{e}$ with a weak probe field. A strong control field is applied between levels $\ket{e}$ and $\ket{r}$  which allows the formation of ladder system. The vee configuration is formed by coupling levels $\ket{g}$ and $\ket{s}$ with a strong switching field. The detunings of respective fields are defined as $\Delta_{p}=\omega_p - \omega_{eg}$, $\Delta_{c}=\omega_c - \omega_{re}$ and $\Delta_{s}=\omega_s - \omega_{sg}$,  and their Rabi frequencies and wavelengths are denoted by $\Omega$ and  $\lambda$.

For specificity, we have considered the energy levels of rubidium atom---level $\ket{g}$ is the $\rm{5S}_{1/2}$ ground state; levels $\ket{e}$ and $\ket{s}$ are $5\rm{P}_{3/2}$ and $5\rm{P}_{1/2}$ excited states; and level $\ket{r}$ is the $44\rm{D}_{5/2}$ Rydberg state. In this case, the wavelengths of applied fields are: $\lambda_p=780$ nm, $\lambda_{s}=795$ nm and $\lambda_{c}=480$ nm. Also, the decay rates of these states are: $\Gamma_g=0$, $\Gamma_e/2\pi=6.1$ MHz, $\Gamma_s/2\pi=5.9$ MHz and $\Gamma_r/2\pi=0.3$ MHz. Since the probe field is weak, it ensures that levels $\ket{e}$ and $\ket{r}$ are always unpopulated and population cycles between levels $\ket{g}$ and $\ket{s}$. This level scheme has been utilized for the realization of EIA using a Rydberg state \cite{BWN16}. 
		
Using the density matrix formalism, the optical Bloch equations---by incorporating the decay of atoms from each level and repopulation from excited levels---for the populations are written by
\begin{equation}
	\begin{aligned}
	\dot{\rho}_{gg}&=\Gamma_{e}\rho_{ee}+\Gamma_{s}\rho_{ss}+\frac{i}{2}\Omega_{p}\left(\rho_{eg}-\rho_{ge}\right)
	+\frac{i}{2}\Omega_{s}\left(\rho_{sg}-\rho_{gs}\right)\\
	\dot{\rho}_{ee}&=-\Gamma_{e}\rho_{ee}+\Gamma_{r}\rho_{rr}+\frac{i}{2}\Omega_{p}\left(\rho_{ge}-\rho_{eg}\right)
	+\frac{i}{2}\Omega_{c}\left(\rho_{re}-\rho_{er}\right)\\
	\dot{\rho}_{ss}&=-\Gamma_{s}\rho_{ss}+\frac{i}{2}\Omega_{s}\left(\rho_{gs}-\rho_{sg}\right)\\
	\dot{\rho}_{rr}&=-\Gamma_{r}\rho_{rr}+\frac{i}{2}\Omega_{c}\left(\rho_{er}-\rho_{re}\right)\\
	\label{population_eqns}
	\end{aligned}
\end{equation} 	
and the equations for the coherences can be written as follows
\begin{equation}
	\begin{aligned}
	\dot{\rho}_{eg}&=\gamma_{eg}\rho_{eg}-\frac{i}{2}\Omega_{p}\left(\rho_{ee}-\rho_{gg}\right)-\frac{i}{2}\left(\Omega_{s}\rho_{es}-\Omega_{c}\rho_{rg}\right)\\
	\dot{\rho}_{sg}&=\gamma_{sg}\rho_{sg}-\frac{i}{2}\Omega_{s}\left(\rho_{ss}-\rho_{gg}\right)-\frac{i}{2}\Omega_{p}\rho_{se}\\
	\dot{\rho}_{rg}&=\gamma_{rg}\rho_{rg}+\frac{i}{2}\left(\Omega_{c}\rho_{eg}-\Omega_{p}\rho_{re}-\Omega_{s}\rho_{rs}\right)\\
	\dot{\rho}_{se}&=\gamma_{se}\rho_{se}+\frac{i}{2}\left(\Omega_{s}\rho_{ge}-\Omega_{p}\rho_{sg}-\Omega_{c}\rho_{sr}\right)\\
	\dot{\rho}_{re}&=\gamma_{re}\rho_{re}-\frac{i}{2}\Omega_{c}\left(\rho_{rr}-\rho_{ee}\right)-\frac{i}{2}\Omega_{p}\rho_{rg}\\
	\dot{\rho}_{rs}&=\gamma_{rs}\rho_{rs}+\frac{i}{2}\left(\Omega_{c}\rho_{es}-\Omega_{s}\rho_{rg}\right)\\
\label{coherence_eqns}
\end{aligned}
\end{equation} 
The $\gamma$'s are defined as
\begin{equation*}
\begin{aligned}
	\gamma_{eg}&=\displaystyle{-\frac{\Gamma_{e}}{2}+i\Delta_{p}}; \,\,\, 
	\gamma_{sg}=\displaystyle{-\frac{\Gamma_{s}}{2}+i\Delta_{s}}; \,\,\, 
	\gamma_{rg}=\displaystyle{-\frac{\Gamma_{r}}{2}+i(\Delta_{p}+\Delta_{c})} ; \\
	\gamma_{se}&=\displaystyle{-\frac{\Gamma_{s}+\Gamma_{e}}{2}+i(\Delta_{s}-\Delta_{p})}; \,\,\, 
	\gamma_{re}=\displaystyle{-\frac{\Gamma_{r}+\Gamma_{e}}{2}+i\Delta_{c}}; \\
	\gamma_{rs}&=\displaystyle{-\frac{\Gamma_{r}+\Gamma_{s}}{2}+i(\Delta_{p}+\Delta_{c}-\Delta_{s})}.
\label{gammas}
\end{aligned}
\end{equation*}

The steady state solution of coupled density matrix-equations for $\rho_{eg}$ under weak probe conditions is given as
\begin{equation}
	\begin{aligned}
		\rho_{eg}&=-\frac{i\Omega_{p}\rho_{gg}}{2\gamma_{eg}\beta} 
		+\frac{i\Omega_{p}\Omega_{s}^2(\rho_{gg}-\rho_{ss})}{8\gamma_{eg}\gamma_{gs}\gamma_{es}\beta} \left(1-\frac{\Omega_{c}^2}{4\gamma_{es}\gamma_{rs}\alpha}-
		\frac{\Omega_{c}^2}{4\gamma_{rg}\gamma_{rs}\alpha}\right)
		\label{solution}
	\end{aligned}
\end{equation}
where
\begin{equation*}
	\begin{aligned}
		&\rho_{gg}-\rho_{ss} = \left[ 1 + \frac{\Omega_{s}^2}{2\left(\frac{\Gamma_{s}^2}{4}+\Delta_{s}^2\right)} \right]^{-1} \,\\
		& \beta=1+\displaystyle{\frac{\Omega_{s}^2}{4\gamma_{eg}\gamma_{es}}+\frac{\Omega_{c}^2}{4\gamma_{eg}\gamma_{rg}}
			-\frac{\Omega_{s}^2 \, \Omega_{c}^2}{16\gamma_{eg}\gamma_{rs}\alpha}\left[\frac{1}{\gamma_{es}}+\frac{1}{\gamma_{rg}}\right]^2} \,\\ 
		&\alpha=\displaystyle{1+\frac{\Omega_{s}^2}{4\gamma_{rg}\gamma_{rs}}+\frac{\Omega_{c}^2}{4\gamma_{es}\gamma_{rs}}} \,\\ 
	\end{aligned}
\end{equation*}

In the present work, the observable is the response of atoms to the weak probe field. The dispersion ($\eta$) and absorption ($\mathcal{A}$) of probe field is proportional to real and imaginary parts of $\rho_{eg}$, and are given by \cite{VAN96}
\begin{equation}
\begin{aligned}
\eta={\Re}\left\{\dfrac{\rho_{eg}\Gamma_e}{\Omega_p}\right\} \,\, {\rm{and}} \,\,
\mathcal{A}={\Im}\left\{ \dfrac{\rho_{eg}\Gamma_e}{\Omega_p}\right\}  
\label{disab}
\end{aligned}
\end{equation}

\subsection{Group index and group velocity}
The group index of the probe field can be evaluated in terms of susceptibility ($\chi$) through the relation
\begin{equation}
n_{g}=1+ \dfrac{\Re\{\chi\}}{2}+ \dfrac{\omega_{p}}{2}\dfrac{\partial \Re\{\chi\}}{\partial\omega_{p}}
\label{groupindex}
\end{equation}
where
\begin{equation*}
	\chi=\displaystyle{\frac{N|\mu_{eg}|^2}{\hbar\varepsilon_{0}\Omega_{p}}}\rho_{eg}
	\label{susceptability}
\end{equation*}
and corresponding group velocity is
\begin{equation}
v_g=\dfrac{c}{n_g}
\label{groupvelocity}
\end{equation}
Here, $c$ is the velocity of light in vacuum, $\omega_p$ is probe frequency, $N$ is the atom number density in the medium, and $|\mu_{eg}|$ is the magnitude of dipole matrix element between levels $\ket{e}$ and $\ket{g}$. The above equation shows that $v_g$ is inversely proportional to the slope of dispersion curve. Large group index ($n_g \gg  1$) due to steep positive dispersion leads to sub-luminal  ($v_g \ll c$) propagation. On the other hand, small ($n_g < 1$) or negative group index due to negative dispersion shows super-luminal ($v_g > c$) propagation.

\subsection{Doppler averaging}
For an atomic vapor at room temperature, we have to account for thermal velocity distribution by carrying out Doppler averaging, which is the typical condition used for an experimental realization. We consider an atom with velocity $v$ interacting with co-propagating probe and switching fields, and a counter-propagating control field \cite{BHN15,BWN16}. Since the wavelengths of probe and control fields are mismatched, the two photon absorption for the ladder sub-system is non-Doppler free, i.e.\ $k_p v - k_{c} v \neq 0$, and probe absorption spreads over different velocity classes \cite{BWN16}. 

We perform Doppler averaging using the one-dimensional Maxwell-Boltzmann velocity distribution. For an atom of mass $M$ at temperature $T$, the Maxwell-Boltzmann distribution of velocities is given by
\begin{equation}
	f(v)dv	= \sqrt{\dfrac{M}{2 \pi k_B T}} \exp{\left(-\dfrac{Mv^2}{2 k_B T}\right)}dv
	\label{MBdistribution}
\end{equation}
where $k_B$ is the Boltzmann constant.

\section{Results and discussion}
\subsection{Sub- and super-luminal propagation}
\subsubsection{Stationary atoms}
\begin{figure}
	\centering
	\includegraphics[width=.4\textwidth]{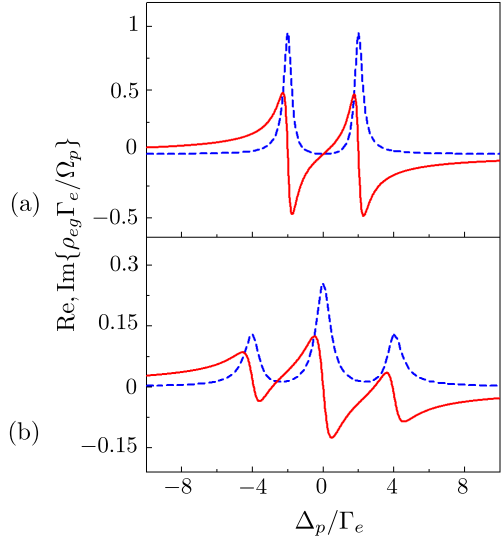}
	\caption{(Color online) Probe dispersion and absorption given by $\Re\{\rho_{eg}\Gamma_e /\Omega_p$\} (solid) and $\Im\{\rho_{eg}\Gamma_e /\Omega_p$\} (dashed) for stationary atoms. In this figure, $\Delta_c =\Delta_s = 0$, and (a) $\Omega_c = 4 \Gamma_e$, $\Omega_s = 0$ and (b) $\Omega_c = \Omega_s = 4 \Gamma_e$.} 
	\label{disabsstationary}
\end{figure}

We first consider the results for stationary atoms. The dispersion (solid curve) and absorption (dashed curve) of weak probe field as a function of probe detuning---using Eqns.\ \eqref{solution} and \eqref{disab}---are shown in Fig.\ \ref{disabsstationary}. Part (a) shows the dispersion profile for Rydberg EIT (switching field is absent) and part (b) shows the dispersion profile for Rydberg EIA (switching field is present). For the case of Rydberg EIT, dispersion profile has positive slope near zero detuning, which can therefore be used for sub-luminal light propagation. On the other hand, dispersion profile for Rydberg EIA has negative slope at line center, which leads to super-luminal light propagation. These results shows that the slope of dispersion profile changes from positive to negative in the presence of switching field. This leads to the switching of group velocity of the probe field from sub- to super-luminal domain.
\begin{figure}
 	\centering
 	\includegraphics[width=.4\textwidth]{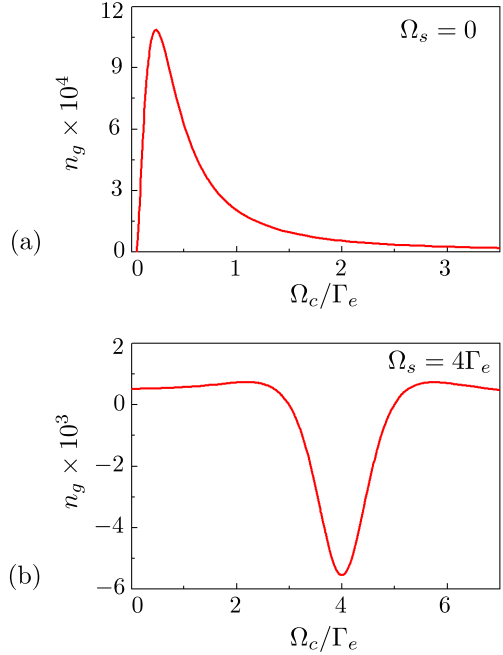}
 	\caption{(Color online) Variation of group index with control field's Rabi frequency, (a) $\Omega_s=0$ and  (b) $\Omega_s=4\Gamma_e$. In this figure $N = 1\times 10^{10}$ atoms/cm$^3$, $\Omega_p= 0.16 \times 10^{-3} \Gamma_e$, and $\Delta_p=\Delta_c=\Delta_s=0$.} 
 	\label{ngstationary}
\end{figure}

Now we discuss the dependence of sub- and super-luminal behavior on the strengths of control and switching fields. We present the variation of $n_g$ versus control Rabi frequency by using Eq.\ \eqref{groupindex}. The results for stationary atoms are shown in Fig.\ \ref{ngstationary}. In the absence of switching field [Fig.\ \ref{ngstationary}(a)], the group index for Rydberg EIT first increases and then decreases with control Rabi frequency, but always corresponds to sub-luminal propagation. In the presence of switching field [Fig.\ \ref{ngstationary}(b)], the large positive $n_g$ becomes negative, which is the condition for Rydberg EIA. The negative $n_g$ again becomes positive with further increase in control Rabi frequency. We thus obtain a \textit{valley of negative $n_g$} in the region of enhanced absorption. Thus an atomic medium can be tuned from sub- to super-luminal propagation by using a four-level vee+ladder configuration. 

\subsubsection{Doppler averaging at room temperature}
\begin{figure}
	\centering
	\includegraphics[width=.4\textwidth]{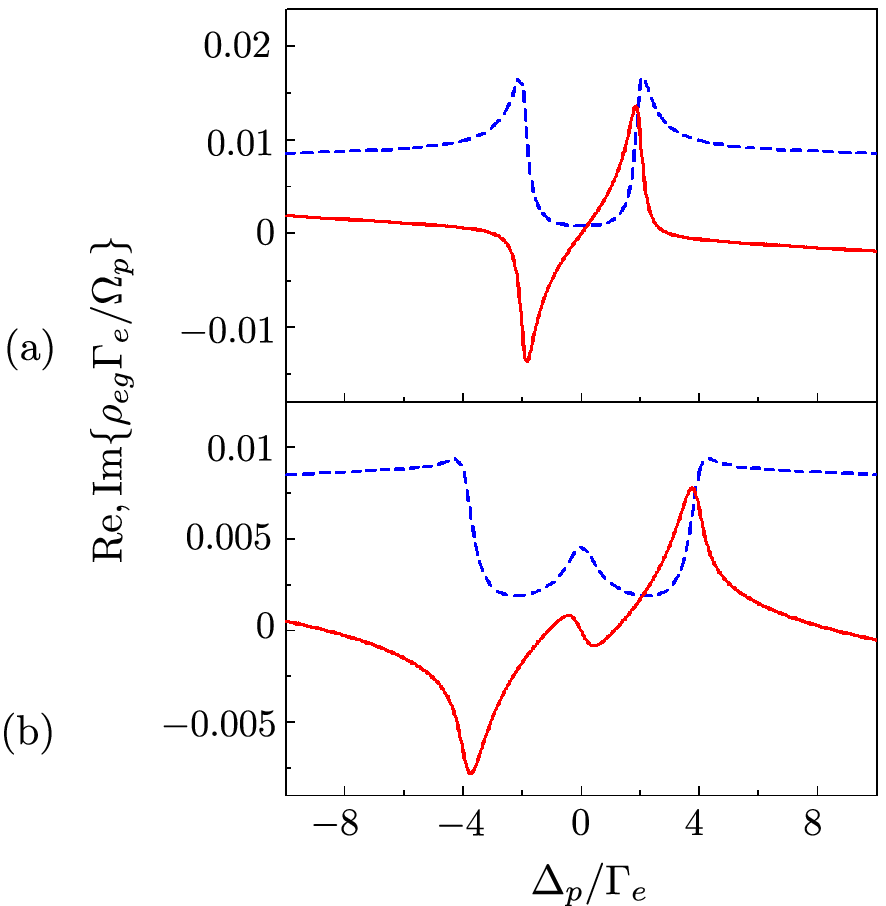}
	\caption{(Color online) Probe dispersion and absorption given by $\Re\{\rho_{eg}\Gamma_e /\Omega_p$\} (solid) and $\Im\{\rho_{eg}\Gamma_e /\Omega_p$\} (dashed) after carrying Doppler averaging at room temperature. All the parameters are same as in Fig.\ \ref{disabsstationary}.}
	\label{disabsthavg}
\end{figure}

We now present the results for Doppler averaging for moving atoms by using Maxwell-Boltzmann velocity distribution given in Eq.\ \eqref{MBdistribution}. The averaging is performed by covering a velocity range of $-500$ to $+500$ m/s, which is adequate for room temperature atoms. The corresponding dispersion and absorption profiles are shown in Fig.\ \ref{disabsthavg}. For the case of Rydberg EIT [Fig.\ \ref{disabsthavg}(a)], the transparency window with positive slope of dispersion curve at line center does not get narrower, due to mismatched wavelengths of probe and control fields \cite{KUS09,BHW14}. In addition, the EIA peak with negative slope is less prominent [Fig.\ \ref{disabsthavg}(b)] again due to mismatched wavelengths \cite{BWN16}. 

\begin{figure}
	\centering
	\includegraphics[width=.4\textwidth]{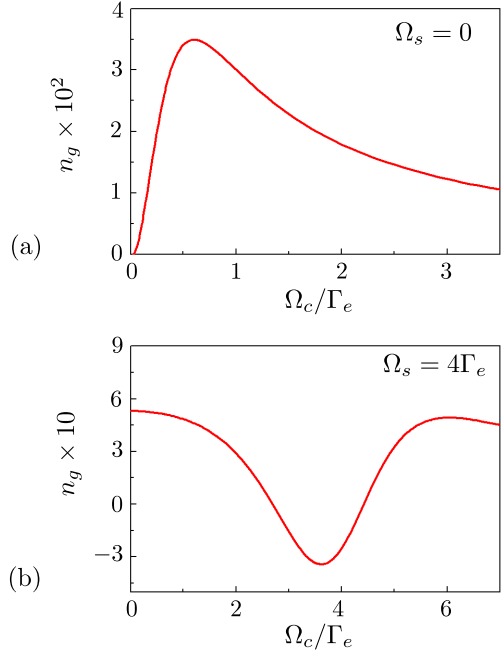}
	\caption{(Color online) Variation of group index with control field's Rabi frequency after carrying out Doppler averaging at room temperature, (a) $\Omega_s=0$ and  (b) $\Omega_s=4\Gamma_e$. All other parameters are same as Fig.\ \ref{ngstationary}.} 
	\label{ngthavg}
\end{figure} 

Now we discuss the effect of atomic velocity on group index. The results at room temperature after carrying out Doppler averaging are shown in Fig.\ \ref{ngthavg}. The values of $n_g$ are reduced by few orders as compared with the results for stationary atoms shown in Fig.\ \ref{ngstationary}. For a 3-level ladder system shown in part (a), the usual sub-luminal propagation is observed. On the other hand, in a 4-level system shown in part (b), the \textit{valley of negative $n_g$} and switching from sub- to super-luminal propagation is observed. The minimum value is slightly shifted due to mismatched wavelengths of the three fields.

\subsection{Transient behavior}
\begin{figure}
	\centering
	\includegraphics[width=.4\textwidth]{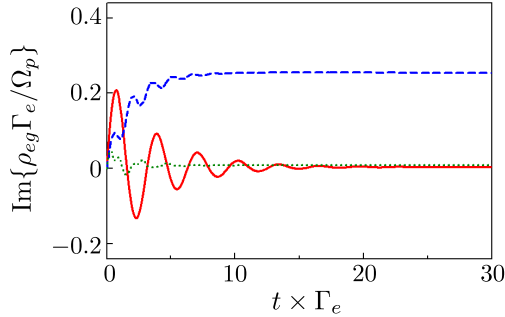}
	\caption{(Color online) Time evolution of $\Im\{\rho_{eg}\Gamma_e/\Omega_p\}$: solid curve is for $\Omega_c=4\Gamma_e,\Omega_s=0$, dashed curve is for $\Omega_c=\Omega_s=4\Gamma_e$ and dotted curve is for $\Omega_c=4\Gamma_e,\Omega_s= 10\Gamma_e$.} 
	\label{transient}
\end{figure}

We now discuss the transient properties of 4-level vee+ladder scheme. The  consideration of the transient behavior of an atomic system is important  due  to  its  potential applications such as absorptive optical switching \cite{LIX95}, in which the transmission of a highly absorptive medium is controlled by  an  additional field. We solve time dependent density matrix equations numerically to show time evolution of $\Im\{\rho_{eg}\Gamma_e/\Omega_p\}$ in Fig.\ \ref{transient}.
In the absence of switching field (solid curve), the damped oscillations become steady and probe absorption reduces to zero \cite{LIX95}. Now, if the switching field is turned on---with Rabi frequency equal to control field---the enhanced probe absorption is observed (dashed curve) \cite{BWN16,SRH04}. With further increase in switching Rabi frequency, the enhanced absorption again reduces close to zero (dotted curve). Thus, a transparent atomic medium can be switched to enhanced absorption and again to reduced absorption by increasing the strength of switching field.  Therefore, the considered 4-level system can be utilized as absorptive switch by turning on switching field \cite{SRH04,HAY98}.

\section{Conclusion}
In summary, we have studied sub- and super-luminal light propagation using a Rydberg state in 4-level vee+ladder rubidium system. We presented results for stationary and moving atoms at room temperature by performing Doppler averaging. The dispersion profile for Rydberg EIT showed positive slope at line center, whereas it is negative for Rydberg EIA obtained when an additional control field is turned on. The variation of group index with control and switching Rabi frequencies showed a \textit{valley of negative group index} that indicates large super-luminality. Therefore, an atomic system consisting of a Rydberg state can be tuned from sub- to super-luminal propagation by applying an additional switching field. We also discussed transient properties of system and showed that the vee+ladder scheme can be utilized as an absorptive optical switch.

\section*{Acknowledgements}
This work was supported by the Department of Science and Technology, India. VB acknowledges financial support from a DS Kothari post-doctoral fellowship of the University Grants Commission, India (grant no. F.4-2/2006 (BSR)/PH/14-15/0069). 


\begin{thebibliography}{29}%
\makeatletter
\providecommand \@ifxundefined [1]{%
 \@ifx{#1\undefined}
}%
\providecommand \@ifnum [1]{%
 \ifnum #1\expandafter \@firstoftwo
 \else \expandafter \@secondoftwo
 \fi
}%
\providecommand \@ifx [1]{%
 \ifx #1\expandafter \@firstoftwo
 \else \expandafter \@secondoftwo
 \fi
}%
\providecommand \natexlab [1]{#1}%
\providecommand \enquote  [1]{``#1''}%
\providecommand \bibnamefont  [1]{#1}%
\providecommand \bibfnamefont [1]{#1}%
\providecommand \citenamefont [1]{#1}%
\providecommand \href@noop [0]{\@secondoftwo}%
\providecommand \href [0]{\begingroup \@sanitize@url \@href}%
\providecommand \@href[1]{\@@startlink{#1}\@@href}%
\providecommand \@@href[1]{\endgroup#1\@@endlink}%
\providecommand \@sanitize@url [0]{\catcode `\\12\catcode `\$12\catcode
  `\&12\catcode `\#12\catcode `\^12\catcode `\_12\catcode `\%12\relax}%
\providecommand \@@startlink[1]{}%
\providecommand \@@endlink[0]{}%
\providecommand \url  [0]{\begingroup\@sanitize@url \@url }%
\providecommand \@url [1]{\endgroup\@href {#1}{\urlprefix }}%
\providecommand \urlprefix  [0]{URL }%
\providecommand \Eprint [0]{\href }%
\providecommand \doibase [0]{http://dx.doi.org/}%
\providecommand \selectlanguage [0]{\@gobble}%
\providecommand \bibinfo  [0]{\@secondoftwo}%
\providecommand \bibfield  [0]{\@secondoftwo}%
\providecommand \translation [1]{[#1]}%
\providecommand \BibitemOpen [0]{}%
\providecommand \bibitemStop [0]{}%
\providecommand \bibitemNoStop [0]{.\EOS\space}%
\providecommand \EOS [0]{\spacefactor3000\relax}%
\providecommand \BibitemShut  [1]{\csname bibitem#1\endcsname}%
\let\auto@bib@innerbib\@empty
\bibitem [{\citenamefont {Boller}\ \emph {et~al.}(1991)\citenamefont {Boller},
  \citenamefont {Imamo\ifmmode~\breve{g}\else \u{g}\fi{}lu},\ and\
  \citenamefont {Harris}}]{BIH91}%
  \BibitemOpen
  \bibfield  {author} {\bibinfo {author} {\bibfnamefont {K.-J.}\ \bibnamefont
  {Boller}}, \bibinfo {author} {\bibfnamefont {A.}~\bibnamefont
  {Imamo\ifmmode~\breve{g}\else \u{g}\fi{}lu}}, \ and\ \bibinfo {author}
  {\bibfnamefont {S.~E.}\ \bibnamefont {Harris}},\ }\href {\doibase
  10.1103/PhysRevLett.66.2593} {\bibfield  {journal} {\bibinfo  {journal}
  {Phys. Rev. Lett.}\ }\textbf {\bibinfo {volume} {66}},\ \bibinfo {pages}
  {2593} (\bibinfo {year} {1991})}\BibitemShut {NoStop}%
\bibitem [{\citenamefont {Harris}(1997)}]{HAR97}%
  \BibitemOpen
  \bibfield  {author} {\bibinfo {author} {\bibfnamefont {S.~E.}\ \bibnamefont
  {Harris}},\ }\href@noop {} {\bibfield  {journal} {\bibinfo  {journal} {Phys.
  Today}\ }\textbf {\bibinfo {volume} {50}},\ \bibinfo {pages} {36} (\bibinfo
  {year} {1997})}\BibitemShut {NoStop}%
\bibitem [{\citenamefont {Fleischhauer}\ \emph {et~al.}(2005)\citenamefont
  {Fleischhauer}, \citenamefont {Imamoglu},\ and\ \citenamefont
  {Marangos}}]{FIM05}%
  \BibitemOpen
  \bibfield  {author} {\bibinfo {author} {\bibfnamefont {M.}~\bibnamefont
  {Fleischhauer}}, \bibinfo {author} {\bibfnamefont {A.}~\bibnamefont
  {Imamoglu}}, \ and\ \bibinfo {author} {\bibfnamefont {J.~P.}\ \bibnamefont
  {Marangos}},\ }\href {\doibase 10.1103/RevModPhys.77.633} {\bibfield
  {journal} {\bibinfo  {journal} {Rev. Mod. Phys.}\ }\textbf {\bibinfo {volume}
  {77}},\ \bibinfo {eid} {633} (\bibinfo {year} {2005})}\BibitemShut {NoStop}%
\bibitem [{\citenamefont {Mohapatra}\ \emph {et~al.}(2007)\citenamefont
  {Mohapatra}, \citenamefont {Jackson},\ and\ \citenamefont {Adams}}]{MJA07}%
  \BibitemOpen
  \bibfield  {author} {\bibinfo {author} {\bibfnamefont {A.~K.}\ \bibnamefont
  {Mohapatra}}, \bibinfo {author} {\bibfnamefont {T.~R.}\ \bibnamefont
  {Jackson}}, \ and\ \bibinfo {author} {\bibfnamefont {C.~S.}\ \bibnamefont
  {Adams}},\ }\href {\doibase 10.1103/PhysRevLett.98.113003} {\bibfield
  {journal} {\bibinfo  {journal} {Phys. Rev. Lett.}\ }\textbf {\bibinfo
  {volume} {98}},\ \bibinfo {pages} {113003} (\bibinfo {year}
  {2007})}\BibitemShut {NoStop}%
\bibitem [{\citenamefont {Tauschinsky}\ \emph {et~al.}(2013)\citenamefont
  {Tauschinsky}, \citenamefont {Newell}, \citenamefont {van Linden van~den
  Heuvell},\ and\ \citenamefont {Spreeuw}}]{TNL13}%
  \BibitemOpen
  \bibfield  {author} {\bibinfo {author} {\bibfnamefont {A.}~\bibnamefont
  {Tauschinsky}}, \bibinfo {author} {\bibfnamefont {R.}~\bibnamefont {Newell}},
  \bibinfo {author} {\bibfnamefont {H.~B.}\ \bibnamefont {van Linden van~den
  Heuvell}}, \ and\ \bibinfo {author} {\bibfnamefont {R.~J.~C.}\ \bibnamefont
  {Spreeuw}},\ }\href {\doibase 10.1103/PhysRevA.87.042522} {\bibfield
  {journal} {\bibinfo  {journal} {Phys. Rev. A}\ }\textbf {\bibinfo {volume}
  {87}},\ \bibinfo {pages} {042522} (\bibinfo {year} {2013})}\BibitemShut
  {NoStop}%
\bibitem [{\citenamefont {Bason}\ \emph {et~al.}(2008)\citenamefont {Bason},
  \citenamefont {Mohapatra}, \citenamefont {Weatherill},\ and\ \citenamefont
  {Adams}}]{BMW08}%
  \BibitemOpen
  \bibfield  {author} {\bibinfo {author} {\bibfnamefont {M.~G.}\ \bibnamefont
  {Bason}}, \bibinfo {author} {\bibfnamefont {A.~K.}\ \bibnamefont
  {Mohapatra}}, \bibinfo {author} {\bibfnamefont {K.~J.}\ \bibnamefont
  {Weatherill}}, \ and\ \bibinfo {author} {\bibfnamefont {C.~S.}\ \bibnamefont
  {Adams}},\ }\href {\doibase 10.1103/PhysRevA.77.032305} {\bibfield  {journal}
  {\bibinfo  {journal} {Phys. Rev. A}\ }\textbf {\bibinfo {volume} {77}},\
  \bibinfo {pages} {032305} (\bibinfo {year} {2008})}\BibitemShut {NoStop}%
\bibitem [{\citenamefont {Kubler}\ \emph {et~al.}(2009)\citenamefont {Kubler},
  \citenamefont {Shaffer}, \citenamefont {Baluktsian}, \citenamefont {Low},\
  and\ \citenamefont {Pfau}}]{KSB09}%
  \BibitemOpen
  \bibfield  {author} {\bibinfo {author} {\bibfnamefont {H.}~\bibnamefont
  {Kubler}}, \bibinfo {author} {\bibfnamefont {J.~P.}\ \bibnamefont {Shaffer}},
  \bibinfo {author} {\bibfnamefont {T.}~\bibnamefont {Baluktsian}}, \bibinfo
  {author} {\bibfnamefont {R.}~\bibnamefont {Low}}, \ and\ \bibinfo {author}
  {\bibfnamefont {T.}~\bibnamefont {Pfau}},\ }\href@noop {} {\bibfield
  {journal} {\bibinfo  {journal} {Nature Photonics}\ }\textbf {\bibinfo
  {volume} {4}},\ \bibinfo {pages} {112} (\bibinfo {year} {2009})}\BibitemShut
  {NoStop}%
\bibitem [{\citenamefont {Tanasittikosol}\ \emph {et~al.}(2011)\citenamefont
  {Tanasittikosol}, \citenamefont {Pritchard}, \citenamefont {Maxwell},
  \citenamefont {Gauguet}, \citenamefont {Weatherill}, \citenamefont
  {Potvliege},\ and\ \citenamefont {Adams}}]{TPM11}%
  \BibitemOpen
  \bibfield  {author} {\bibinfo {author} {\bibfnamefont {M.}~\bibnamefont
  {Tanasittikosol}}, \bibinfo {author} {\bibfnamefont {J.~D.}\ \bibnamefont
  {Pritchard}}, \bibinfo {author} {\bibfnamefont {D.}~\bibnamefont {Maxwell}},
  \bibinfo {author} {\bibfnamefont {A.}~\bibnamefont {Gauguet}}, \bibinfo
  {author} {\bibfnamefont {K.~J.}\ \bibnamefont {Weatherill}}, \bibinfo
  {author} {\bibfnamefont {R.~M.}\ \bibnamefont {Potvliege}}, \ and\ \bibinfo
  {author} {\bibfnamefont {C.~S.}\ \bibnamefont {Adams}},\ }\href
  {http://stacks.iop.org/0953-4075/44/i=18/a=184020} {\bibfield  {journal}
  {\bibinfo  {journal} {Journal of Physics B: Atomic, Molecular and Optical
  Physics}\ }\textbf {\bibinfo {volume} {44}},\ \bibinfo {pages} {184020}
  (\bibinfo {year} {2011})}\BibitemShut {NoStop}%
\bibitem [{\citenamefont {Goren}\ \emph {et~al.}(2004)\citenamefont {Goren},
  \citenamefont {Wilson-Gordon}, \citenamefont {Rosenbluh},\ and\ \citenamefont
  {Friedmann}}]{GWR04}%
  \BibitemOpen
  \bibfield  {author} {\bibinfo {author} {\bibfnamefont {C.}~\bibnamefont
  {Goren}}, \bibinfo {author} {\bibfnamefont {A.~D.}\ \bibnamefont
  {Wilson-Gordon}}, \bibinfo {author} {\bibfnamefont {M.}~\bibnamefont
  {Rosenbluh}}, \ and\ \bibinfo {author} {\bibfnamefont {H.}~\bibnamefont
  {Friedmann}},\ }\href {\doibase 10.1103/PhysRevA.69.053818} {\bibfield
  {journal} {\bibinfo  {journal} {Phys. Rev. A}\ }\textbf {\bibinfo {volume}
  {69}},\ \bibinfo {pages} {053818} (\bibinfo {year} {2004})}\BibitemShut
  {NoStop}%
\bibitem [{\citenamefont {Bason}\ \emph {et~al.}(2009)\citenamefont {Bason},
  \citenamefont {Mohapatra}, \citenamefont {Weatherill},\ and\ \citenamefont
  {Adams}}]{BMW09}%
  \BibitemOpen
  \bibfield  {author} {\bibinfo {author} {\bibfnamefont {M.~G.}\ \bibnamefont
  {Bason}}, \bibinfo {author} {\bibfnamefont {A.~K.}\ \bibnamefont
  {Mohapatra}}, \bibinfo {author} {\bibfnamefont {K.~J.}\ \bibnamefont
  {Weatherill}}, \ and\ \bibinfo {author} {\bibfnamefont {C.~S.}\ \bibnamefont
  {Adams}},\ }\href {http://stacks.iop.org/0953-4075/42/i=7/a=075503}
  {\bibfield  {journal} {\bibinfo  {journal} {Journal of Physics B: Atomic,
  Molecular and Optical Physics}\ }\textbf {\bibinfo {volume} {42}},\ \bibinfo
  {pages} {075503} (\bibinfo {year} {2009})}\BibitemShut {NoStop}%
\bibitem [{\citenamefont {Chanu}\ \emph {et~al.}(2012)\citenamefont {Chanu},
  \citenamefont {Pandey},\ and\ \citenamefont {Natarajan}}]{CPN12}%
  \BibitemOpen
  \bibfield  {author} {\bibinfo {author} {\bibfnamefont {S.~R.}\ \bibnamefont
  {Chanu}}, \bibinfo {author} {\bibfnamefont {K.}~\bibnamefont {Pandey}}, \
  and\ \bibinfo {author} {\bibfnamefont {V.}~\bibnamefont {Natarajan}},\ }\href
  {http://stacks.iop.org/0295-5075/98/i=4/a=44009} {\bibfield  {journal}
  {\bibinfo  {journal} {EPL (Europhysics Letters)}\ }\textbf {\bibinfo {volume}
  {98}},\ \bibinfo {pages} {44009} (\bibinfo {year} {2012})}\BibitemShut
  {NoStop}%
\bibitem [{\citenamefont {Bharti}\ and\ \citenamefont {Wasan}(2013)}]{BHW13}%
  \BibitemOpen
  \bibfield  {author} {\bibinfo {author} {\bibfnamefont {V.}~\bibnamefont
  {Bharti}}\ and\ \bibinfo {author} {\bibfnamefont {A.}~\bibnamefont {Wasan}},\
  }\href {http://stacks.iop.org/0953-4075/46/i=12/a=125501} {\bibfield
  {journal} {\bibinfo  {journal} {J. Phys. B}\ }\textbf {\bibinfo {volume}
  {46}},\ \bibinfo {pages} {125501} (\bibinfo {year} {2013})}\BibitemShut
  {NoStop}%
\bibitem [{\citenamefont {Bharti}\ and\ \citenamefont
  {Natarajan}(2015)}]{BHN15}%
  \BibitemOpen
  \bibfield  {author} {\bibinfo {author} {\bibfnamefont {V.}~\bibnamefont
  {Bharti}}\ and\ \bibinfo {author} {\bibfnamefont {V.}~\bibnamefont
  {Natarajan}},\ }\href {\doibase
  http://dx.doi.org/10.1016/j.optcom.2015.08.042} {\bibfield  {journal}
  {\bibinfo  {journal} {Optics Communications}\ }\textbf {\bibinfo {volume}
  {356}},\ \bibinfo {pages} {510 } (\bibinfo {year} {2015})}\BibitemShut
  {NoStop}%
\bibitem [{\citenamefont {Bharti}\ \emph {et~al.}(2016)\citenamefont {Bharti},
  \citenamefont {Wasan},\ and\ \citenamefont {Natarajan}}]{BWN16}%
  \BibitemOpen
  \bibfield  {author} {\bibinfo {author} {\bibfnamefont {V.}~\bibnamefont
  {Bharti}}, \bibinfo {author} {\bibfnamefont {A.}~\bibnamefont {Wasan}}, \
  and\ \bibinfo {author} {\bibfnamefont {V.}~\bibnamefont {Natarajan}},\ }\href
  {\doibase http://dx.doi.org/10.1016/j.physleta.2016.05.038} {\bibfield
  {journal} {\bibinfo  {journal} {Physics Letters A}\ }\textbf {\bibinfo
  {volume} {380}},\ \bibinfo {pages} {2390 } (\bibinfo {year}
  {2016})}\BibitemShut {NoStop}%
\bibitem [{\citenamefont {Hau}\ \emph {et~al.}(1999)\citenamefont {Hau},
  \citenamefont {Harris}, \citenamefont {Dutton},\ and\ \citenamefont
  {Behroozi}}]{HHD99}%
  \BibitemOpen
  \bibfield  {author} {\bibinfo {author} {\bibfnamefont {L.~V.}\ \bibnamefont
  {Hau}}, \bibinfo {author} {\bibfnamefont {S.}~\bibnamefont {Harris}},
  \bibinfo {author} {\bibfnamefont {Z.}~\bibnamefont {Dutton}}, \ and\ \bibinfo
  {author} {\bibfnamefont {C.~H.}\ \bibnamefont {Behroozi}},\ }\href {\doibase
  10.1038/17561} {\bibfield  {journal} {\bibinfo  {journal} {Nature}\ }\textbf
  {\bibinfo {volume} {397}},\ \bibinfo {pages} {594} (\bibinfo {year}
  {1999})}\BibitemShut {NoStop}%
\bibitem [{\citenamefont {Kash}\ \emph {et~al.}(1999)\citenamefont {Kash},
  \citenamefont {Sautenkov}, \citenamefont {Zibrov}, \citenamefont {Hollberg},
  \citenamefont {Welch}, \citenamefont {Lukin}, \citenamefont {Rostovtsev},
  \citenamefont {Fry},\ and\ \citenamefont {Scully}}]{KSZ99}%
  \BibitemOpen
  \bibfield  {author} {\bibinfo {author} {\bibfnamefont {M.~M.}\ \bibnamefont
  {Kash}}, \bibinfo {author} {\bibfnamefont {V.~A.}\ \bibnamefont {Sautenkov}},
  \bibinfo {author} {\bibfnamefont {A.~S.}\ \bibnamefont {Zibrov}}, \bibinfo
  {author} {\bibfnamefont {L.}~\bibnamefont {Hollberg}}, \bibinfo {author}
  {\bibfnamefont {G.~R.}\ \bibnamefont {Welch}}, \bibinfo {author}
  {\bibfnamefont {M.~D.}\ \bibnamefont {Lukin}}, \bibinfo {author}
  {\bibfnamefont {Y.}~\bibnamefont {Rostovtsev}}, \bibinfo {author}
  {\bibfnamefont {E.~S.}\ \bibnamefont {Fry}}, \ and\ \bibinfo {author}
  {\bibfnamefont {M.~O.}\ \bibnamefont {Scully}},\ }\href {\doibase
  10.1103/PhysRevLett.82.5229} {\bibfield  {journal} {\bibinfo  {journal}
  {Phys. Rev. Lett.}\ }\textbf {\bibinfo {volume} {82}},\ \bibinfo {pages}
  {5229} (\bibinfo {year} {1999})}\BibitemShut {NoStop}%
\bibitem [{\citenamefont {Kocharovskaya}\ \emph {et~al.}(2001)\citenamefont
  {Kocharovskaya}, \citenamefont {Rostovtsev},\ and\ \citenamefont
  {Scully}}]{KRS01}%
  \BibitemOpen
  \bibfield  {author} {\bibinfo {author} {\bibfnamefont {O.}~\bibnamefont
  {Kocharovskaya}}, \bibinfo {author} {\bibfnamefont {Y.}~\bibnamefont
  {Rostovtsev}}, \ and\ \bibinfo {author} {\bibfnamefont {M.~O.}\ \bibnamefont
  {Scully}},\ }\href {\doibase 10.1103/PhysRevLett.86.628} {\bibfield
  {journal} {\bibinfo  {journal} {Phys. Rev. Lett.}\ }\textbf {\bibinfo
  {volume} {86}},\ \bibinfo {pages} {628} (\bibinfo {year} {2001})}\BibitemShut
  {NoStop}%
\bibitem [{\citenamefont {Wang}\ \emph {et~al.}(2000)\citenamefont {Wang},
  \citenamefont {Kuzmich},\ and\ \citenamefont {Dogariu}}]{WKD00}%
  \BibitemOpen
  \bibfield  {author} {\bibinfo {author} {\bibfnamefont {L.~J.}\ \bibnamefont
  {Wang}}, \bibinfo {author} {\bibfnamefont {A.}~\bibnamefont {Kuzmich}}, \
  and\ \bibinfo {author} {\bibfnamefont {A.}~\bibnamefont {Dogariu}},\ }\href
  {\doibase 10.1038/35018520} {\bibfield  {journal} {\bibinfo  {journal}
  {Nature}\ }\textbf {\bibinfo {volume} {406}},\ \bibinfo {pages} {277}
  (\bibinfo {year} {2000})}\BibitemShut {NoStop}%
\bibitem [{\citenamefont {Dogariu}\ \emph {et~al.}(2001)\citenamefont
  {Dogariu}, \citenamefont {Kuzmich},\ and\ \citenamefont {Wang}}]{DKW01}%
  \BibitemOpen
  \bibfield  {author} {\bibinfo {author} {\bibfnamefont {A.}~\bibnamefont
  {Dogariu}}, \bibinfo {author} {\bibfnamefont {A.}~\bibnamefont {Kuzmich}}, \
  and\ \bibinfo {author} {\bibfnamefont {L.~J.}\ \bibnamefont {Wang}},\ }\href
  {\doibase 10.1103/PhysRevA.63.053806} {\bibfield  {journal} {\bibinfo
  {journal} {Phys. Rev. A}\ }\textbf {\bibinfo {volume} {63}},\ \bibinfo
  {pages} {053806} (\bibinfo {year} {2001})}\BibitemShut {NoStop}%
\bibitem [{\citenamefont {Agarwal}\ and\ \citenamefont
  {Dasgupta}(2004)}]{AGD04}%
  \BibitemOpen
  \bibfield  {author} {\bibinfo {author} {\bibfnamefont {G.~S.}\ \bibnamefont
  {Agarwal}}\ and\ \bibinfo {author} {\bibfnamefont {S.}~\bibnamefont
  {Dasgupta}},\ }\href {\doibase 10.1103/PhysRevA.70.023802} {\bibfield
  {journal} {\bibinfo  {journal} {Phys. Rev. A}\ }\textbf {\bibinfo {volume}
  {70}},\ \bibinfo {pages} {023802} (\bibinfo {year} {2004})}\BibitemShut
  {NoStop}%
\bibitem [{\citenamefont {Han}\ \emph {et~al.}(2005)\citenamefont {Han},
  \citenamefont {Guo}, \citenamefont {Bai},\ and\ \citenamefont {Sun}}]{HGB05}%
  \BibitemOpen
  \bibfield  {author} {\bibinfo {author} {\bibfnamefont {D.}~\bibnamefont
  {Han}}, \bibinfo {author} {\bibfnamefont {H.}~\bibnamefont {Guo}}, \bibinfo
  {author} {\bibfnamefont {Y.}~\bibnamefont {Bai}}, \ and\ \bibinfo {author}
  {\bibfnamefont {H.}~\bibnamefont {Sun}},\ }\href {\doibase
  http://dx.doi.org/10.1016/j.physleta.2004.11.022} {\bibfield  {journal}
  {\bibinfo  {journal} {Physics Letters A}\ }\textbf {\bibinfo {volume}
  {334}},\ \bibinfo {pages} {243 } (\bibinfo {year} {2005})}\BibitemShut
  {NoStop}%
\bibitem [{\citenamefont {qi~Kuang}\ \emph {et~al.}(2009)\citenamefont
  {qi~Kuang}, \citenamefont {gang Wan}, \citenamefont {Kou}, \citenamefont
  {Jiang},\ and\ \citenamefont {yue Gao}}]{KWK09}%
  \BibitemOpen
  \bibfield  {author} {\bibinfo {author} {\bibfnamefont {S.}~\bibnamefont
  {qi~Kuang}}, \bibinfo {author} {\bibfnamefont {R.}~\bibnamefont {gang Wan}},
  \bibinfo {author} {\bibfnamefont {J.}~\bibnamefont {Kou}}, \bibinfo {author}
  {\bibfnamefont {Y.}~\bibnamefont {Jiang}}, \ and\ \bibinfo {author}
  {\bibfnamefont {J.}~\bibnamefont {yue Gao}},\ }\href {\doibase
  10.1364/JOSAB.26.002256} {\bibfield  {journal} {\bibinfo  {journal} {J. Opt.
  Soc. Am. B}\ }\textbf {\bibinfo {volume} {26}},\ \bibinfo {pages} {2256}
  (\bibinfo {year} {2009})}\BibitemShut {NoStop}%
\bibitem [{\citenamefont {Kaur}\ \emph {et~al.}(2014)\citenamefont {Kaur},
  \citenamefont {Bharti},\ and\ \citenamefont {Wasan}}]{KBW14}%
  \BibitemOpen
  \bibfield  {author} {\bibinfo {author} {\bibfnamefont {P.}~\bibnamefont
  {Kaur}}, \bibinfo {author} {\bibfnamefont {V.}~\bibnamefont {Bharti}}, \ and\
  \bibinfo {author} {\bibfnamefont {A.}~\bibnamefont {Wasan}},\ }\href
  {\doibase 10.1080/09500340.2014.931479} {\bibfield  {journal} {\bibinfo
  {journal} {Journal of Modern Optics}\ }\textbf {\bibinfo {volume} {61}},\
  \bibinfo {pages} {1339} (\bibinfo {year} {2014})}\BibitemShut {NoStop}%
\bibitem [{\citenamefont {Vemuri}\ \emph {et~al.}(1996)\citenamefont {Vemuri},
  \citenamefont {Agarwal},\ and\ \citenamefont {Nageswara~Rao}}]{VAN96}%
  \BibitemOpen
  \bibfield  {author} {\bibinfo {author} {\bibfnamefont {G.}~\bibnamefont
  {Vemuri}}, \bibinfo {author} {\bibfnamefont {G.~S.}\ \bibnamefont {Agarwal}},
  \ and\ \bibinfo {author} {\bibfnamefont {B.~D.}\ \bibnamefont
  {Nageswara~Rao}},\ }\href {\doibase 10.1103/PhysRevA.53.2842} {\bibfield
  {journal} {\bibinfo  {journal} {Phys. Rev. A}\ }\textbf {\bibinfo {volume}
  {53}},\ \bibinfo {pages} {2842} (\bibinfo {year} {1996})}\BibitemShut
  {NoStop}%
\bibitem [{\citenamefont {Kumar}\ and\ \citenamefont {Singh}(2009)}]{KUS09}%
  \BibitemOpen
  \bibfield  {author} {\bibinfo {author} {\bibfnamefont {M.~A.}\ \bibnamefont
  {Kumar}}\ and\ \bibinfo {author} {\bibfnamefont {S.}~\bibnamefont {Singh}},\
  }\href {\doibase 10.1103/PhysRevA.79.063821} {\bibfield  {journal} {\bibinfo
  {journal} {Phys. Rev. A}\ }\textbf {\bibinfo {volume} {79}},\ \bibinfo
  {pages} {063821} (\bibinfo {year} {2009})}\BibitemShut {NoStop}%
\bibitem [{\citenamefont {Bharti}\ and\ \citenamefont {Wasan}(2014)}]{BHW14}%
  \BibitemOpen
  \bibfield  {author} {\bibinfo {author} {\bibfnamefont {V.}~\bibnamefont
  {Bharti}}\ and\ \bibinfo {author} {\bibfnamefont {A.}~\bibnamefont {Wasan}},\
  }\href {\doibase http://dx.doi.org/10.1016/j.optcom.2014.03.057} {\bibfield
  {journal} {\bibinfo  {journal} {Optics Communications}\ }\textbf {\bibinfo
  {volume} {324}},\ \bibinfo {pages} {238} (\bibinfo {year}
  {2014})}\BibitemShut {NoStop}%
\bibitem [{\citenamefont {qing Li}\ and\ \citenamefont {Xiao}(1995)}]{LIX95}%
  \BibitemOpen
  \bibfield  {author} {\bibinfo {author} {\bibfnamefont {Y.}~\bibnamefont {qing
  Li}}\ and\ \bibinfo {author} {\bibfnamefont {M.}~\bibnamefont {Xiao}},\
  }\href {\doibase 10.1364/OL.20.001489} {\bibfield  {journal} {\bibinfo
  {journal} {Opt. Lett.}\ }\textbf {\bibinfo {volume} {20}},\ \bibinfo {pages}
  {1489} (\bibinfo {year} {1995})}\BibitemShut {NoStop}%
\bibitem [{\citenamefont {Shen}\ \emph {et~al.}(2004)\citenamefont {Shen},
  \citenamefont {Ruan},\ and\ \citenamefont {He}}]{SRH04}%
  \BibitemOpen
  \bibfield  {author} {\bibinfo {author} {\bibfnamefont {J.-Q.}\ \bibnamefont
  {Shen}}, \bibinfo {author} {\bibfnamefont {Z.-C.}\ \bibnamefont {Ruan}}, \
  and\ \bibinfo {author} {\bibfnamefont {S.}~\bibnamefont {He}},\ }\href
  {\doibase http://dx.doi.org/10.1016/j.physleta.2004.07.074} {\bibfield
  {journal} {\bibinfo  {journal} {Physics Letters A}\ }\textbf {\bibinfo
  {volume} {330}},\ \bibinfo {pages} {487 } (\bibinfo {year}
  {2004})}\BibitemShut {NoStop}%
\bibitem [{\citenamefont {Harris}\ and\ \citenamefont
  {Yamamoto}(1998)}]{HAY98}%
  \BibitemOpen
  \bibfield  {author} {\bibinfo {author} {\bibfnamefont {S.~E.}\ \bibnamefont
  {Harris}}\ and\ \bibinfo {author} {\bibfnamefont {Y.}~\bibnamefont
  {Yamamoto}},\ }\href {\doibase 10.1103/PhysRevLett.81.3611} {\bibfield
  {journal} {\bibinfo  {journal} {Phys. Rev. Lett.}\ }\textbf {\bibinfo
  {volume} {81}},\ \bibinfo {pages} {3611} (\bibinfo {year}
  {1998})}\BibitemShut {NoStop}%
\end{thebibliography}

%

\end{document}